# Intergenerational transmission of culture among immigrants: Gender gap in education among first and second generations *

*By* Hamid NOGHANIBEHAMBARI ᵃ†
Nahid TAVASSOLI ᵇ & Farzaneh NOGHANI ᶜ

**Abstract.** This paper illustrates the intergenerational transmission of the gender gap in education among first and second-generation immigrants. Using the Current Population Survey (1994-2018), we find that the difference in female-male education persists from the home country to the new environment. A one standard deviation increase of the ancestral country's female-male difference in schooling is associated with 17.2% and 2.5% of a standard deviation increase in the gender gap among first and second generations, respectively. Since gender perspective in education uncovers a new channel for cultural transmission among families, we interpret the findings as evidence of cultural persistence among first generations and partial cultural assimilation of second generations. Moreover, Disaggregation into country-groups reveals different paths for this transmission: descendants of immigrants of lower-income countries show fewer attachments to the gender opinions of their home country. Average local education of natives can facilitate the acculturation process. Immigrants residing in states with higher education reveal a lower tendency to follow their home country attitudes regarding the gender gap.

## 1. Introduction

T he issues regarding immigrants and their assimilation into the host countries have been among the main concerns among policymakers. These issues have been reflected in the US policy debates recently and have raised new questions regarding integration and isolation of immigrants. This fact can be better viewed by the demographics of immigrants in the US population as well as the steady rise in the inflow of immigrants. Roughly 24 percent of the US population consists of first and second-generation immigrants and the authorized resident permits ᵃ † Department of Economics, Texas Tech University, Lubbock, TX, 79409, USA, USA.
☎ . 806-620-1812.  ✉ . Hamid.NoghaniBehambari@ttu.edu
ᵇ Department of Economics, Texas Tech University, Lubbock, TX, 79409, USA, USA.
☎ . 806-519-3929.  ✉ . nahid.tavassoli@ttu.edu
ᶜ Jerry S. Rawls College of Business, Administration Building, Texas Tech University, Flint Ave, Lubbock, TX, USA.
☎ . 806-252-3094.  ✉ . farzaneh.noghani@ttu.edu

experienced an average growth rate of 1.3 percent during the last two decades (Security, 2018; Trevelyan *et al.,* 2016).

A relatively wide literature in economics documents that some cultural characteristics are inherited through the intergenerational transmission process (Black *et al.,* 2005; Doepke & Zilibotti, 2017; Noghanibehambari *et al.,* 2020; Tomes, 1981). Looking at the intergenerational correlation between parents' and children's characteristics suffer from Endogeneity issues since they are exposed to some common environmental and political factors. Immigration provides a semi-natural experiment to isolate these common factors. These institutional factors are eliminated when we look at immigrants coming from different countries but are now exposed to the same economic and political system. By isolating the institutional elements, one can look into the correlation between home-country characteristics and behaviors of a different generation of immigrants in order to explore the potential intergenerational transmission of cultural traits (Alesina *et al.,* 2013).

Gender perspective could potentially capture some facets of culture. In presence of intergenerational transmission, descendants of countries in which more traditional views towards women exist are expected to continue such views in the host countries. Hence, female education, female labor force participation, and fertility rates are proxied for culture by some recent papers (Blau *et al.,* 2013; Nollenberger *et al.,* 2016). Moreover, a decrease in the explanatory power of home-country characteristics suggests that immigrants' behavior is going under structural changes, segregating from the path of intergenerational inheritance of traits, and dissolving into the new environment.

This paper explores a new channel for cultural transmission among first and second-generation immigrants using Current Population Data (CPS) covering the years 1994-2018. We proxy culture by educational attainments and later by gender difference in education. As a new innovation to add to this literature, we narrow down the cultural proxy by assuming that, ceteris paribus, a family's perspective towards culture can be decomposed into a pure belief about investment in human capital and a gender-discriminative belief towards the distribution of this investment. We proxy the former aspect by levels of education and confirm the intergenerational transmission which is consistent with current literature. Later, we proxy the latter aspect of culture by the gender gap in educational attainments inthe country of ancestry. I, then, proceed to search for the persistence of such gender differences among first and second-generation immigrants. Education has been used in the literature as one proper proxy for culture. However, if households value education differently for their male offspring than their female children, and so behave discriminatory in their educational spending for their female and male offspring, the gender gap in education can capture some gender-biased opinions and thus reveals a cultural trait.

We find that a one standard deviation increase across home-countries in educational-difference is associated with 17.2% of a standard deviation increase in gender difference in education among first generations. Moreover, a one standard deviation increase in the educational gender gap across fathers' homeland leads to 2.5% of a standard deviation increase in gender difference in education among second generations.

In addition, we disaggregate the data into two country groups: low-income and high-income countries. Descendants of low-income countries are less likely to follow the path of their home-country gender-gap in education. A one standard deviation increase in the father's birthplace gender-difference in education will increase the gender gap of second generations by 2.6% of a standard deviation for high-income groups while insignificant for low-income home countries. Comparing both groups within a regression shows a significantly lower persistence of the gender gap among low-income second generations compared to descendants of high-income countries.

Contextual factors can accelerate the assimilation process and facilitate closing the gender gap. We show that state average education and state average education of natives could significantly lower the attachments of immigrants to their home country gender gap in education. Moreover, the marginal impacts of native education on the gender gap is higher and more effective for the first generations compared to the second generations. After all, we provide some insights on the possible explanations of lower attachment of immigrants from a low-income country to their home country aspects, both levels and gender difference in education, compared to descendants of high-income countries.

As a final analysis to capture the gender gap in education, we narrow down the definition of education from years of schooling to college graduates in degrees of Art and Humanities, a field that has been suggested by sociology literature to be the basis of cultural capital (DiMaggio & Mukhtar, 2004). A gender difference in these majors reflects the gender bias of families in investment in the cultural capital of their children. We show that a 10% difference in degrees awarded in the fields of Art and Humanities in the country of ancestry is associated with a 1% increase in gender difference of awarded degrees among immigrants, and this coefficient is quite robust among first and second generations and over different specifications. This fact reveals that there is some gender bias in attaining cultural capital which is transmitted intergenerationally.

This paper makes a number of contributions to the current literature. Taking advantage of some cross-country data on education instead of constructing a separate data set based on past Census data (prevalent in the literature) boosts the reliability of results besides the fact that it makes a significantly larger dataset that covers considerably more countries. Such a larger number of source countries in the dataset allows me to examine different specification checks at country-group levels. We disaggregate the data for two country-groups, based on income per capita, which gives me an opportunity to seek different transmission pace between low-income and high-income countries. This vision of country-group difference in cultural transmission for second-generation immigrants in the US has not been analyzed by current waves of research. Blau & Kahn (2007) highlights that different aspect of culture take several generations to dissolve in the new environment while adaptation and assimilation of some facets are much faster. If so, and assuming a family's orientation towards education can be thought of independently as their gender gap perspectives, we construct a model to assess intergenerational transmission of gender disparities among immigrants by using the gender gap in education as the relevant proxy. Thus, compared to the ongoing literature, we break up the cultural proxy and show different levels of transmission for each aspect. Furthermore, instead of using the same variables and characteristics of the home country for all first and second-generation immigrants, we allow for different cohorts to be affected by different cohort-specific gender-specific variables of their source country. In addition, investigating the intergenerational transmission of cultural capital and gender difference in cultural capital, proxied by degrees awarded in Art and Humanities qualifications, is a novel contribution to this literature.

The rest of the paper is organized as follows: section 0 reviews some related papers to this study. In section 0, we describe the core data sets, sample selection strategy, and give a brief summary statistics of my final dataset. Basic results to capture the levels of home country education on immigrants' outcomes conditional and unconditional on peer effects are analyzed in section 0. In section 0, we construct a distinct model to investigate for intergenerational transmission of gender difference in education conditional and unconditional on individual characteristics. Section 0 focuses on intergenerational transmission of gender difference in cultural capital proxied by degrees in Art and Humanities. Some discussions on identification strategy and its possible drawbacks are provided in section 0. Ultimately, we impart some concluding remarks in section 0.

## 2. Literature review

Immigration provides a semi-natural experiment to study the impact of social norms on economic and non-economic outcomes. Being exposed to the same political and institutional factors in the host country, immigrants vary in their economic and non-economic outcomes. Since culture evolves slowly from one generation to the next, these variations can be partly explained by their cultural traits which had been passed from their home countries to the first and second generations. Therefore, a model imposed on immigrants' outcomes could isolate cultural traits by controlling for political and institutional factors and hence be implemented to assess the intergenerational transmission of prevailing social beliefs. However, due to some reasons such as work permits for new arrival immigrants, citizenship barriers, and language proficiency deficiencies.¹ Therefore, a study on second generations seems more promising since such restraints are expected not to bind for them.²

Fernández (2010) provides a review of recent research on the cultural elements and social factors in determining economic outcomes of different generations of immigrants. As defined in that paper "*The epidemiological approach studies the variation in outcomes across different immigrant groups residing in the same country.*"

Using General Social Survey between the years 1977-87, Fernández & Fogli (2006) examine the role of culture in fertility rates of second (and higher) generations by taking total fertility rates of the country of ancestry at 1950 as a proxy for culture. Moreover, since a woman's own family experience can have an effect on her own fertility, they control for women's own number of siblings, and by doing do they isolate more their cultural proxy. The effects remain significant even when they use different decades for the proxy variable.

Blau *et al.* (2013) revisit Female education, labor supply, and fertility for second-generation American immigrants. ³ They construct the corresponding proxy measures using the behavior of first generations in 1970-2000 Census data as the proxy for their home country characteristics. Applying a similar epidemiological approach using the 1995-2011 March Current Population Survey, they find significant evidence of cultural transmissions among second generations while highlighting the asymmetric effects of father and mother home country characteristics. The effect of fertility from the mother's country of ancestry is larger than that of the father's but the father's sourcecountry education has larger effects for the education of second generations than that of the mother's.

The study of Dustmann *et al.* (2012) compares the gaps in test scores between natives and second generations in some OECD countries and shows that the gaps vary widely across countries and reduce or vanish once conditioning on parental characteristics. It also searches for peer and school quality effects on educational achievements of descendants of Turkish immigrants in several host countries by comparing their test scores with those Turkish children whose parents belong to the same cohorts as those of immigrants but decided not to emigrate. Conditional and unconditional on parental characteristics, a better school system and higher peer quality lead to higher test scores for children of immigrants.

---

¹ Association between language proficiency and employment is investigated by Dustmann & Fabbri (2003). It finds evidence that language expertise of non-white immigrants in UK has a significantly positive effect on likelihood of employment and lack of such competences is associated with earning losses.

² However, as shown by Casey & Dustmann (2008), there are evidences lead to intergenerational transmission of language capital and English fluency. Immigrants' offspring are highly affected by their parents' proficiency conditional on parental and family characteristics.

³ For some studies on education refer to (Anderson, 2014; Bahrs & Schumann, 2019; Barr & Turner, 2015; Dennison, 2019; Noghanibehambari, 2020; NoghaniBehambari *et al.,* 2020a, 2020b).

Issues in the gender pay gap, gender difference in labor supply, and gender gap in human capital investment have previously demanded a large strand of literature (Blau & Kahn, 1997; Cobb-Clark & Moschion, 2017; Goldin, 2014; Goldin *et al.*, 2006; Lavy & Sand, 2015). However, only recently surfaced a new wave of literature to assess the role of culture in explaining the gender gap in labor supply and education. Antecol (2000) uses gender gap in LFPR among home country groups to explain differences in labor supply of first and second-generation immigrants in the 1990 Census. He finds that controlling for observed human capital and individual characteristics, home country differences can explain half of gender gap in labor supply among first generations implying that a portable factor as culture plays a role. However, for second and higher generations such role becomes smaller suggesting some degrees of assimilation.

While the latter paper implements a differential of male and female working hours, Blau *et al.* (2011) focuses separately on each gender's labor supply. It highlights that since male labor force participation is not affected by female labor supply in the home country, the significant coefficients on women's regressions can be attributed to the notion of the gender gap.

Studies on gender disparities in educational achievements have mostly focused on test scores and early educational attainments (Cobb-Clark & Moschion, 2017; Cornwell *et al.*, 2013; Fryer & Levitt, 2010). Implementing the same epidemiological approach, Nollenberger *et al.* (2016) investigate whether the gender gap in math test scores among second-generation immigrants can be explained by gender gaps in their home countries. In order to find a proper proxy for the cultural gender gap, they construct a Gender Gap Index which is a score between 0 and 1 and reflects the economic, political opportunities, education, and well-being of women. They use standardized math test scores of second generations extracted from Program for International Student Assessment data set (2003-2012) in 9 host countries and show that a one standard deviation increase in the gender equality index in their home country causes a 42 percent of a standard deviation reduction in the math gender gap. Moreover, in the case of the math gender gap, the cultural transmission accounts for about two-thirds of all gender-related factors which contribute to the gap.

### 3. Data overview and sample selection

In order to analyze different aspects of the intergenerational transmission of culture, we use Current Population Survey files from January 1994 (the first year that CPS starts to ask mother and father birthplace) to November 2018 extracted from Flood *et al.* (2018). To avoid any double-counting, we eliminated all waves in which it is the second or more times that an individual is participating in the survey.

An immigrant is defined as one who has been born in a foreign country including all 16 U.S territories. A second generation is one who has been born inside the United States but has at least one parent born in a foreign country including all 16 U.S territories. Consistent with the literature, we excluded women below 25, the common threshold age to complete education. All unmatched observations or those for whom data for BPL (Birth Place), MBPL (Mother's Birth Place), or FBPL (Father's Birth Place) is not available are eliminated. All countries with less than 50 observations for either male or female in each analysis in the CPS files are excluded as well.

Since those immigrants who came into the US below the age of 18 and experienced pre-college schools in the US are expected to have experienced a different assimilation trajectory compared to those first generations who came after this age, we excluded all foreign-born individuals whose date of reported entry to the US is 18 years less than their birth year. All in all, the sample consists of 322,786 first-generation individuals and 222,450 second-generation with 106 countries of ancestry.

Panel information on international education is withdrawn from Barro & Lee (2001).[4] They build up the dataset for populations over 15 and 25 years old for male, female, and total population based on the years of schooling from 1950-2000 in five-year intervals. The years of schooling varies from 0 to 17. The data contains 142 countries with at least one observation. There are 107 countries with complete information between the specified intervals among which 106 are matched with CPS and have enough observations in CPS (i.e. more than 50 matched persons in either gender). Moreover, we excluded the 15-year-old group and eliminated total variables since the gender-specific and gender-difference effects are variables of interest in this paper. This dataset is then combined with CPS based on BPL for first generations, and MBPL and FBPL for second generations.

While home country educations are the basis of cultural transmission, the education of white natives is set to serve as a basis for assessing the cultural assimilation and also the gender gap. We compute the average education of white natives at metropolitan area level using a combined dataset from decennial censuses 1990-2000 and annual American Community Survey 2000-2016 extracted from Ruggles *et al.* (2017). We implemented integrated IPUMS person sampling weights. To amalgamate this dataset with CPS, we use a linear extrapolation using 1990 and 2000 census files for missing years in CPS that is 1994-1999.

Table 1 a summary statistics of the final data set is illustrated. The variable *years of schooling* is constructed from detailed codes of immigrants and natives data sets to be in line with home country data sets. Since the coding system of CPS and census is not the same and moreover the Barro-Lee *years of schooling* is up to 17, we report the normalized variables in the lower section of table 1. Average male natives are 0.3% more educated than their female counterparts while among first and second generations this disadvantage is 1.5% and 1.6%, respectively. The last four columns cover home country levels of education for an illustrative year, 1990. On average males above age 25 have attained roughly 12.3% and 6.9% more education than females in low-income and high-income countries, accordingly. The difference for all subgroups is negative implying a disadvantage for females. This difference, however, is less severe for natives compared to immigrants and even compared to the subgroup of high-income countries.

---

[4] The dataset is publicly available as Barro-Lee educational attainment dataset at [Retrieved from].

Figure 2 and Figure 3 show a simple unconditional correlation between the normalized educational attainments of different generations of immigrants, their corresponding home country characteristics, and their native counterparts. The top two panels of Figure 2 demonstrate how education as a cultural proxy has been transmitted from the birth country to the first generations while no such effect is seen with second generations and their mothers' country of ancestry. This fact is reversed when we compare the two generations with their fellow natives. There is a positive (unconditional) correlation between second generations' education and white natives while almost no relation is observed for the first generations and natives. The gender gap in education reveals the same results as shown in Figure 3. The education difference among second generations is more correlated with natives rather than their mother birthplace. The disparities in educational attainments of first generations can be explained by the characteristics of their country of ancestry rather than gender difference among natives.

## 4. The empirical model and main results

4.1. Education as a cultural proxy

In this section, we attempt to capture intergenerational transmission of education among different generations of immigrants in the U.S. using 1994-2018 CPS files implementing the following equation:

$$y_{itsch}^g = \beta_0 + \beta_1 Z_i^g + \beta_2 \bar{y}_{is,native}^g + \alpha_c^g X_{ich}^g + \mu_t + \zeta_h + \lambda_s + \nu_{itsch} \tag{1}$$

In this formulation $y_{itsch}^g$ is the outcome variable for individual $i$ at time $t$ who resides at Metropolitan Statistical Area $s$ with country of ancestry (BPL for first generations, and MBPL or FBPL for second generations) $c$ who belongs to gender group (male or female) $g$ and cohort $h$ for whom the $X$ is the gender-specific cohort-adjusted home country variable. We include some individual characteristics in $Z$ which includes average total family income[5], the number of own siblings[6], and fourth polynomial function for age[7]. It also includes state-year averages of the unemployment rate, family income, share of first and second generations in the state population, and gender inequality index at home country.[8] Average outcomes in the country of ancestry $c$ for gender group $g$ and cohort $h$ is included in the vector $X$, and $\alpha$ is the coefficient of interest that determines the degree of transmission of cultural proxies from source country to different generations of immigrants. $\mu$ is a set of dummies for the year each person attended the CPS survey. In $\lambda$ is included a set of dummies of Statistical Metropolitan Areas. Finally, $v$ is a disturbance term.

---

[5] As suggested by Mayer (1997), family income can be considered as a sign of ability rather than pointing to their educational outcomes which can make a channel to affect children educational outcomes regardless of parentage educational levels.

Some structural changes in immigrants' origin and composition have been documented in the literature which was mostly due to The United States Immigration Act of 1965. The share of European or English language speakers had experienced a dramatic fall from 46 percent in the 1960s to 13 percent in the 1980s (LaLonde & Topel, 1991). New immigrants brought new human skills and distinct social capital which led to a secular change in the quality of new cohorts that has been the standpoint of several studies (Borjas, 1985, 2015). To account for such differences in cohort quality, we categorized first generations into six cohorts based on their reported year of immigration[9]: arrivals before 1960, between 1960-1970, 1971-1980, 1981-1990, 1991-2000, and 2001-2018. Second generations are bracketed into four cohorts based on their age: birth cohorts before 1965, between 1966-1975, 1976-1985, and 1986-1995. Accordingly, four decade-groups of home characteristics (i.e. the years 1950, 1960, 1970, and 1980) are assigned to each cohort. $\zeta$ is a set of cohort dummies to control for cohort fixed effects.

The average outcomes of natives, who belong to the same age range and living in the same metropolitan area, provide a specific path for assimilation that is unlikely to be correlated with immigrants' characteristics and unobserved heterogeneity of individuals. In the case of education, it potentially captures not only peer effect and peer quality, but also school quality and educational system features. In the above specification, $\bar{y}$ is the average education of natives at time $t$ who reside at metropolitan statistical area $s$ with gender $g$.

In equation 1 the parameter $\alpha$ captures the gender-specific channel through which source country average educational achievements affect corresponding variables of first and second generations of immigrants in the United States.

The main results of equation 1 are depicted in Table 2. In columns 7 and 8 results of a full specification regression for female and male second generations are reported. State-year-cohort-specific average education of natives is controlled for in order to take into account the social norms of the new environment. A one standard deviation increase in female's educational attainments across FBPLs increases the female education of second generations by 5.5% of a standard deviation.

---

[6] For discussions on child quantity-quality trade off and more specifically effects of family size on children's educational outcomes refer to (Angrist *et al.*, 2010; Dayiouglu *et al.*, 2009).

[7] The results are not, however, sensitive to including a third degree polynomial of age.

[8] This variable is extracted from Human Development Reports of The United Nations Development Program. The data is publicly available at [Retrieved from].

[9] Lubotsky (2007) explains how the self-reported year of entry to US could refer to the last date of arrival for some transient immigrants and how such differences could lead to severe biases in repeated cross-sectional data to assess the assimilation of immigrants.

Moreover, A one standard deviation increase in males' education across FBPLs is associated with a 6.6% increase of a standard deviation in male' education of second generations. The coefficients are much larger for first generations recorded in the full specification regressions in columns 13 and 14. A one standard deviation increase in female' education inthe home country leads to 38.2% of a standard deviation raise in females' education among first-generation immigrants. Average education of natives, on the other hand, plays a more significant role for second generations. A one standard deviation increase in average education of natives in the Metropolitan statistical area of residence will increase education of second generations by 10.8% and 12.9%, for males and females respectively.

Table 3 and Table 4 split the sample based on two country of ancestry income-groups. First-generation immigrants of low-income countries in the US tend to close the gap with their native counterparts and show less tendency to follow their ancestors' educational levels compared to those from high-income countries. A one standard deviation increase in BPL's female education is associated with a 46.3% and 49.6% increase in women's education among the first generations who were born in low-income and high-income countries respectively. In columns 9 and 10, we check whether the coefficients of the two groups are different. The interaction of an indicator for the low-income country and BPL education is negative and statistically significant. This suggests that first-generation immigrants from low-income countries are less attached to their home country's educational levels compared to immigrants from high-income countries.

These tendencies are mitigated for second generations of both groups (columns 1-2 and 5-6 in Table 3, for females and males, respectively). One standard deviation across FBPLs female education increases the education of second-generation females by 5.5% and 6.4% for low-income and high-income countries, respectively. As shown by the interaction term in column 9, the difference between the coefficients of the two groups is insignificant. On the other hand, the average education of natives has strong explanatory power for descendants of low-income countries implying some assimilation in their educational level (columns 3-4 in Table 3, for females and males, respectively). A one standard deviation increase in average education of natives will increase second-generation females' education of low and high-income countries by 11.5% and 12.7% of a standard deviation, accordingly.

The big picture uncovered by Table 3 and Table 4 is the faster pace of assimilation for second generations originating in countries in lower ladders of income per capita and also lower attachments of the first generations in low-income countries to the characteristics of their country of birth. The United States is itself among high-income countries and so if there is a correlation in educational attainments of countries in this group, then there is less space to be filled by immigrants of high-income countries while for descendants of low-income countries there are an abundance of opportunities to seek in order to fill the gap. Moreover, the average education of low-income countries is much lower than the average education in high-income countries. A second-generation coming from one of these countries might be encouraged to surpass their parents' educational attainments more easily because of better educational systems, easiness in the accessibility of education, higher quality of the schools, and higher returns of education in the labor market. Therefore, he or she is more likely, meanwhile more easily, to enter high school or college. This fact leads to a higher pace of assimilation as shown by the inverse of the coefficients of the educational levels in their mothers' or fathers' birthplace.

Going back to Table 2, positive and highly significant coefficients in all specifications for both first and second-generation immigrants highlight the role of source country in explaining variation in outcomes of both first and second generations. These results are consistent with Blau *et al.* (2013) who used educational attainments of immigrants from 1970-2000 Census data as a proxy for the country of ancestry's educational level. However, the sample size used here is significantly larger, and using a separate dataset on historical average trends in educational level, which had been obtained directly by source country rather than immigrants in the U.S., make the results more reliable.

Had only economic conditions, low skilled demands of jobs in the labor market or lower returns of education in economy driven trends in education in different countries, there would have not been any relation between educational level in the country of ancestry and second generations, once families confront different economic conditions and a labor market with different coordination. However, this fact contradicts the observed results. Once observable factors at individual and household levels are taken into account, an epidemiological approach controls for institutional factors and therefore it leaves us with a residual that shows a high correlation with home country aggregate outcomes. We consider this residual, which is significantly persistent over generations, a cultural drive of individuals' behavior. All in all, The historical origins of cultural formations have been under considerable investigation by recent literature (Alesina *et al.*, 2013; Alesina & Giuliano, 2011; Bisin & Verdier, 2000; Di Tella *et al.*, 2007; Giuliano, 2007; Giuliano & Spilimbergo, 2013).

### 4.2. Gender gap as a cultural proxy

If there are cultural traits that interpret education and gender preferences differently, then our results cannot completely validate cultural transmission or assimilation. However, if cultural solutions regarding educational attainments of individuals act faster than their beliefs about genders, then we must be able to detect lower intergenerational transmission in education for both male and female individuals but with persistent gender inequality. Such gender gap mirrors differentials in social norms and perspectives towards gender roles and so it affects families' distinct investment in their male and female offspring. On the other hand, if a country of ancestry's levels of education has strong positive explanatory power for educational attainments of both genders of second generations, but the gender inequality in the education of their country of ancestry fails to explain the gender differentials in education among second generations, we can claim that families' gender perspectives have gone under structural changes in the new environment. More importantly, the difference between the effects of paternal or maternal source country in education and education inequality validates my decomposition of immigrants' culture that is into two windows: viewpoint regarding education and perspective towards gender role in education.

Firstly, In order to explore, qualitatively, the cultural factor in the educational gender gap, we use cross-country responses to questions regarding gender belief in World Value Survey (WVS) and, later, look into their cross-sectional correlation with gender differences in education. We use waves 3 (1994-1998) and 4 (1999-2004) of the World Value Survey.[10] Weopt for two questions regarding general opinion on gender issues. The first question asks to what extent the respondent agrees with the following statement: "University is more important for boys than girls". The second question asks if, in some circumstances, the respondent is restricted to have only one child would he/she prefer a boy or a girl. Later, we construct a standardized measure of gender-based bias based on these answers and merge the cross-country dataset with a gender-difference measure of years of schooling used in this study. As illustrated in Figure 4 and Figure 5, there is a positive significant correlation between gender-based opinions and gender difference in education. The unconditional correlations between the measures of gender-bias and the gender-difference in education are 39% and 42% for questions one and two, respectively. Thus, we expect that the gender difference in education captures some cultural traits of individuals.

In order to examine this hypothesis, we introduce the following empirical model:

$$y_{itsch}^g = \beta_0 + \beta_1 Z_{isch}^g + \beta_2 S + \beta_3 (X_{ich}^f - X_{ich}^m) + \beta_4 S \times (X_{ich}^f - X_{ich}^m) + \mu_t + \zeta_h + \lambda_s + \nu_{itsch} \qquad (2)$$

Where $S$ is a dummy variable equals 1 if individual $i$ is female and 0 if male. In Z, we include some individual characteristics, which constitute a quadruple polynomial function of age, number of own siblings, and total family income. In this formulation, $\beta_4$ captures the effect of home country gender differential in education on the first and second generations' gender difference in educational attainment. The cohort classification follows the same procedure explicated in section 0.

---

[10] The dataset is publicly available at [Retrieved from].

The main results of the model introduced in equation 2 are reported in Table 5 for both generations of immigrants. In the left section of this table, the full specification results for the gender difference in the education of mother's and father's country of birth are reported in columns 1 and 2. A one standard deviation rise in gender difference in education across fathers' country of ancestry increases the gender gap among second generations by 2.5% of a standard deviation. Note that coefficients on mothers' birthplace are meaningless. At the same time, gender difference among natives shows high explanatory power for second generations. In the full fixed-effect formulation reported in column 5, a one standard deviation change in the educational gender gap among natives is mirrored in gender difference in education among second generations by a change of 11.3% of a standard deviation.

In the right segment of Table 5, the estimations for the first generations are shown. Column 3 reports the effect of the gender gap in the country of birth on the gender gap among first generations for a full specification model of equation 2. Among the first generations, the reflection of home countries' gender difference is almost 8 times more than that of second generations. A one standard deviation increase in educational gender difference in the home country is associated with 19.2% of a standard deviation increase in the gender gap among foreign-born immigrants. Therefore, the gender gap in education, as a proxy for gender-biased opinions, transmit from one generation to the next. However, this cultural transmission is mitigated among second generations implying partial cultural assimilation.

Unobserved heterogeneity among individuals and ethnic groups can bias our coefficients of interests in two ways. As implied by Fernandez & Fogli (2009) if immigrants are different in a systematic way from their counterparts in the country of ancestry then we have a source of bias that is hard to identify and control for. This fact will be aggravated if there are, for example, some criteria for giving visas to applicants or if these kinds of criteria differ from country to country. Secondly, if there are some other cultural traits, like trust, that accelerate or decelerate assimilation and so has some correlation with the cultural aspect of education (or gender difference), it makes the coefficients in equation 1 biased and if different genders absorb these aspects of culture in a systematically different way, then the coefficients in equation 2 will be biased as well. However, recognizing, verifying, and controlling for such traits is arduous and requires much more comprehensive datasets.

As we showed in section 0, the rate of cultural transmission is affected by the initial point. Developing countries, on average, have higher gender differences whereas developed ones have less discrimination towards females, as shown in a preliminary way by summary statistics of education in A1. Tables Table 1. The similarity in the new environment to the old one, as in the case of most high-income countries, lets less space to be filled and based on this less pressure to assimilate. In the opposite direction, more space is provided for descendants of low-income countries who face a much more different environment, usually with more opportunities for women, and a more welcoming atmosphere in the labor market. This difference could encourage more women to participate in educational attainments. Thus, we expect less attachment of educational difference among second generations to the gender difference in the education of their country of ancestry. To capture this facet, the model introduced in equation 2 is run separately for immigrants of different country groups. The results are shown in Table 6 and Table 7 for second and first generations, respectively.

The first column in the left two segments of Table 6 reports the coefficient of FBPL for low and high-income source countries. A one standard deviation change in FBPL's education difference is associated with a 2.6% standard deviation change in the gender gap of second generations of high-income countries while the coefficient is insignificant for low-income regions. The rightmost panel compares the coefficients of two groups within the same regression. The first column shows the interaction of FBPL's gender difference and an indicator for the low-income country. The negative sign implies that descendants of low-income countries are less attached to their ancestral countries' characteristics. This fact is confirmed by the MBPL's gender difference in the second column. Gender difference in education across FBPLs or MBPLs has more explanatory power for second generations of high-income countries compared to those of low-income countries.

The same comparison can be obtained for the first generations in Table 7. A one standard deviation increase in BPL's gender difference in education is reflected in a 14% increase in gender disparity in education among first generations of low-income countries while an 11.7% increase for immigrants of high-income countries (First column in each segment). Again, to compare the coefficients we run a regression for the pooled sample. The interaction term of the low-income home country and gender difference measure is negative and statistically significant. This fact is in line with results of Table 6 and country-group results of section 0. First-generation immigrants from low-income countries reveal lower persistence in their cultural attitudes regarding gender-based opinions.

The results in this section imply that the gender gap in education contains some cultural aspects, namely, gender biasedness of families toward the education of their offspring. Although this gender disparity mitigates for second generations, it remains significant. Moreover, the intergenerational transmission of the gender gap in education is stronger and more persistent for immigrants (both first and second generations) of high-income countries compared to low-income countries suggesting a stronger momentum of gender role beliefs among the former groups. Although only full specifications are reported in Table 6 and Table 7 the coefficients are quite robust conditional and unconditional on individual characteristics and vary only slightly by including or omitting the fixed effects.

Equation 2 takes differences at the community level, i.e. those individuals linked to the same ancestry, in the same year, located in the same area in US. Thus, we expect that all fixed effects at the community level that does not change over time are eliminated by the first difference and should not play a significant role in explaining the variations of gender gap. However, if these fixed effects can have differential impacts on educational level of male and females then we should include them in the model. For example the network of people in a specific metropolitan area who are from the same country could drive parts of educational level of males. Meanwhile, their gender biased opinion could drive the education of females differently. Here, including fixed effects could omit these time invariant effects. Furthermore, we avoid to narrow down the difference in education from local area into household level for two reasons. First, restricting the dataset into families with more than one child (second generations) who reside with their families even after completing the years of education (after 25 years of age) will result in a very small sample size which restrains the income-group analysis. Second, and more noticeably, the purpose of the identification strategy in equation 2 is to capture families' gender-based opinions. Using a family fixed effect model will, more probably, eliminate this factor in the first difference.

Table 6 and Table 7 provide evidence that the assimilation process depends on the initial home country characteristic, i.e. GNI per capita. However, assimilation also depends on contextual factors such as local-area-specific education, income, and the share of immigrants in the population. The higher concentration of immigrants could slow down the native-immigrant gap in gender opinions. On the other hand, the higher educated local population might have lower gender biased opinions and so facilitate the process of closing the gender gap among immigrants who come from countries with higher gender differences in education. In order to check the effect of contextual factors in the assimilation process, we interact state average characteristics with $\Delta(X)$, the aggregate gender difference at home country, in equation 2. The year-specific state characteristics are withdrawn from the same CPS files and include average family income, average education, average education of white Native Americans, and share of first and second-generation immigrants in the population. The results are reported in Table 8 and Table 9 for first and second generations. The education of natives plays a significant role in the acculturation process of both generations. As shown in columns 5 and 6 in Table 9, the state level year-specific average education of Native Americans could significantly lower the attachment of second generations to the gender difference of MBPL and FBPL, respectively. Gender differences among second generations who reside in states with higher educational levels are less affected by their country of ancestry's gender gap in education compared to those who reside in states with lower education. However, the interaction term with other state covariates, namely income and share of second-generation immigrants (columns 1-4), are insignificant. Therefore, the educational level of natives in the state is the main contextual catalyzer to speed up the assimilation process in case of the educational gender gap.

## 5. Gender gap in degrees awarded

In this section, we narrow down the definition of education from the broad measure of years of schooling to a specific degree field that is more likely to be correlated with culture. Sociologists consider the fields of Arts and Humanities the foundations of cultural capital (DiMaggio & Mukhtar, 2004). If families in different societies have different cultural capital which is partly because of their aggregate investment in these fields, then moving backward, we can isolate more the cultural proxy by using degrees awarded in the area of Arts and Humanities. Moreover, we can use the gender gap in degrees awarded in these areas of study in order to capture the gender disparity of different societies in their cultural capital. We apply equation 2 and a linear probability model to investigate whether gender differences in the degrees awarded in Art and Humanities are transmitted intergenerationally to the first and second-generation immigrants. In this analysis, the dependent variable is a dummy equals one if the individual has a degree in the related majors and zero otherwise.

I use the distribution of tertiary degrees awarded in humanities and art qualifications by sex in OECD countries inthe year 2017.[11] The difference between female and male percentage in the awarded degrees is used as home country proxies.

The CPS does not ask about the respondent's field of study. An alternative dataset is the American Community Survey which has two advantages for our analysis. First, it asks for detailed information about the individual's field of study or the degree awarded. Second, in addition to the large sample sizes which allows more immigrants to be identified, also it asks a broad set of questions on labor supply and earnings. One drawback is that while there is information on the country of birth and their ancestry there is no direct measure of the birth country of mother or father. Thus, it is not possible to distinguish between second and higher generations. The significance and consistency of the coefficients of interest in all specifications and for different generations imply, however, that this fact should not bring up concerns. We exclude all individuals with unmatched data for the country of origin or missing data for any of the covariates. Moreover, age is restricted to be less than 19, to eliminate all with less than (potential) college-age, and 50, to avoid large disturbances in measures of culture since the OECD awarded degree dataset is for the year 2017. Moreover, we restrict the sample to include all with at least some college degree and for whom there is a reported degree. Finally, there will be 708,559 immigrants from 19 ancestries and 25 countries of birth which satisfy the mentioned restrictions.

---

[11] The dataset is publicly available at [Retrieved from].

The results are reported in Table 10. The average gender difference in the OECD dataset is 27.8% with a standard deviation of 12.6%. In the full specification for first generations (column 3, left panel), a 10% increase in gender difference in the awarded degrees in the home country will increase the gender difference by 1%. Referring to the right panel, a 10% increase in the gender difference in countries of ancestry is associated with a 1% increase in gender difference in degrees awarded in Art and Humanities among second generations. The coefficients are quite robust conditional and unconditional on individual and family characteristics. These results suggest that gender disparities in cultural capital are being transmitted from one generation to the next.

## 6. Discussion

Although the epidemiological approach taken in this study is widely implemented in the literature it has some drawbacks. First, we operate under the assumption that average home country characteristics represent the behavior of all immigrants in their country of origin. However, as the culture is different in different parts of a country so are the social norms that immigrants could have been exposed to in their country of origin. A better approach is to look at the distribution of the cultural variables in different strata of the home country and different immigrants based on their observable demographics or socioeconomic characteristics. This difference in immigrants' features and their source country characteristics is also documented in the literature. For instance, Chiquiar & Hanson (2005) use US and Mexico censuses and show that Mexican immigrants in the US are more educated than those nonimmigrants who decided not to migrate and reside in Mexico.

Second, Assimilation, as an essential facet of immigration literature and policy, comprises distinct dimensions not fully interdependent. As Dustmann (1996) noted, cultural assimilation and economic assimilation could move along parallel rays while being affected by the same factors but with various degrees. It analyzed the feeling of national identity as a proxy for cultural assimilation and finds that personal characteristics and initial nationality affects this feeling of identity while labor outcomes are surprisingly irrelevant. Using 22 waves of German Socio-Economic Panel Data Casey & Dustmann (2010) revisits the intergenerational transmission of identity and finds strong evidence supporting intergenerational transmission of identity while such outcomes are weakly affected by labor supply variables. From an immigrant perspective, Angelini *et al.* (2015) documents a strong association between variables measuring cultural assimilation and the subjective well-being of different generations of immigrants. Hence, not only both sorts of assimilation matter in a policy perspective but also these two facets, cultural assimilation, and economic assimilation, are not necessarily and perfectly intercorrelated. As Casey & Dustmann (2010) showed feeling of identity, as a cultural proxy, is not correlated with labor outcomes and it is more likely to pass from one generation to another than being dissolved so soon. Future works might distinguish among distinct aspects of cultural transmission.

Third, the visa selection could bias the intergenerational links. It could be the case that US visas are granted based on criteria that differ across countries or differ across individuals within a country. If these criteria are correlated with determinants of culture then there isa potential bias in the estimated coefficients. For instance, if more visas are issued to higher educated individuals who, for some unobservable reasons, have lower gender discriminative opinions compared to their peers in their home country, then we should expect coefficients that are under-biased.

## 7. Conclusion

The intergenerational transmission of traits from the home country to the source country is of great importance in immigration policy designs. The issue of whether the gender-based opinions could be assimilated in the new environment or which aspects are more persistent is essential in gender equality policy makings.It is well established that education and gender-biased opinions have some cultural forces that transmit from one generation to the next. This paper provides evidence that gender-based opinions are, partly, reflected in education. The gender gap in education among first and second-generation immigrants can be traced back to their home countries. Female immigrants coming from countries in which females have much lower educational levels compared to males will suffer from a similar disadvantage even after migrating to the US. However, this gap closes partially for second generations.

Next, we split the data into two source country income groups. Immigrants from high-income countries show more persistence in their gender-based opinions. Compared to low-income countries, the coefficients of gender difference in source country are, partly, higher for descendants of high-income countries. For both generations, the gender gap in the low-income home country has lower explanatory power for the gender gap in education among immigrants.

Adding the average gender difference among natives reveals the same trend. The gender gap among immigrants fromlow-income countries is more correlated with the gender gap of white Native Americans while the correlation is slightly higher for immigrants fromlow-income countries.This confirms the fact that gender-discriminatory opinions persist with lower momentum among immigrants of low-income countries and the rates of assimilation, regarding gender biasedness, is much faster among immigrants of these countries.

Why immigrants from lower-income countries have a lower attachment to their home country characteristics, namely education and educational gender gap? Female labor force participation is lower in low-income countries. In the US the schools, colleges, and workplaces are much more diverse than in low-income countries. Such diversity has already been established in high-income countries. Thus, female immigrants from low-income countries face opportunities that were not available for their ancestors. This facet could close the gender gap in education by increasing the rates of school enrolment and college enrolment of women immigrants. Moreover, a labor market with lower gender-biased selection, lower gender-biased colleagues, higher returns to education, and a more welcoming environment for women could encourage females of low-income countries to attend school at much higher rates than their high-income counterparts. These facts can explain the lower attachments of immigrants from low-income countries to their homeland facets.

Next, we investigate the effect of contextual factors in the process of assimilation. State average education of natives (those white Americans with the same age group who reside in the same state in the same year) can significantly lower the attachment of immigrants to their country of ancestry characteristics. For both generations, the gender gap in the home country has a lower effect on the gender difference in the education of those immigrants who reside in states with higher average native education compared to immigrants in states with lower education. However, since the choice of place is endogenous, we avoid interpreting the results as cause and effect relationship. The estimated coefficients on the interaction of contextual factors are only association.

Finally, we narrow down the proxy of culture to a commonly used measure of cultural capital: degrees awarded in majors of Arts and Humanities. Using this new proxy, we investigate whether there is intergenerational transmission of the gender gap in the share of awardees in these majors. We find that immigrants of countries in which cultural capital is more equally distributed among genders have a tendency to invest in the cultural capital of their female and male offspring more equally. The coefficients are quite robust in different specifications and similar in magnitude for both first and second generations.

**Notes**

*Disclosure Statement*: The authors have no conflicts of interest to disclose.
*Funding Details*: The authors received no financial support for the research, authorship, and publication of this article.

# Appendix
## A1. Tables
**Table 1.** *Summary Statistics*

|  | Natives | | Immigrants (CPS) | | Home Country (Barro-Lee) | |
|---|---|---|---|---|---|---|
|  | Census-ACS | CPS | First Generations | Second Generations | Low Income | High Income |
| *Education (Years of Schooling)* | | | | | | |
| Female | 13.22 | 13.43 | 11.81 | 13.24 | 3.86 | 7.56 |
|  | (2.82) | (2.58) | (4.52) | (2.95) | (2.61) | (2.14) |
| Male | 13.25 | 13.50 | 12.12 | 13.58 | 5.47 | 8.33 |
|  | (2.96) | (2.77) | (4.72) | (3.10) | (2.24) | (2.04) |
| Difference (f-m) | -0.025 | -0.070 | -0.320 | -0.334 | -1.61 | -0.77 |
|  | (0.102) | (0.132) | (0.292) | (0.202) | (0.88) | (0.77) |
| *Education (Normalized)* | | | | | | |
| Female | 0.629 | 0.640 | 0.562 | 0.631 | 0.304 | 0.603 |
|  | (0.135) | (0.123) | (0.216) | (0.141) | (0.212) | (0.173) |
| Male | 0.631 | 0.643 | 0.577 | 0.647 | 0.426 | 0.672 |
|  | (0.141) | (0.132) | (0.225) | (0.148) | (0.192) | (0.175) |
| Difference (f-m) | -0.0012 | -0.0033 | -0.0153 | -0.0159 | -0.123 | -0.069 |
|  | (0.0048) | (0.0063) | (0.0139) | (0.0096) | (0.069) | (0.063) |
| Observations | 43,107,416 | 3,161,012 | 322,786 | 222,450 | 54 | 62 |

**Notes.** Standard deviations are in parentheses.

Table 2 - *Regression Analysis for the effects of average years of schooling in the country of ancestry on educational attainments of first and second-generation immigrants*

| | Second Generations | | | | | | | | First Generations | | | | | |
|---|---|---|---|---|---|---|---|---|---|---|---|---|---|---|
| DV: Education | Female b/se | Male b/se | Female b/se | Male b/se | Female b/se | Male b/se | Female b/se | Male b/se | Female b/se | Male b/se | Female b/se | Male b/se | Female b/se | Male b/se |
| Female Educ, MBPL | 0.055 (0.048) | | | | 0.054 (0.047) | | 0.053 (0.034) | | | | | | | |
| Male Educ, FBPL | 0.068*** (0.019) | | | | 0.067*** (0.019) | | 0.055*** (0.012) | | | | | | | |
| Male Educ, FBPL | | 0.077*** (0.022) | | | | 0.077*** (0.022) | | 0.066*** (0.013) | | | | | | |
| Male Educ, MBPL | | 0.062 (0.050) | | | | 0.062 (0.050) | | 0.050 (0.034) | | | | | | |
| Avg Educ, Natives | | | 0.131*** (0.016) | 0.114*** (0.012) | 0.127*** (0.015) | 0.114*** (0.012) | 0.129*** (0.016) | 0.108*** (0.018) | | | 0.084*** (0.017) | 0.052* (0.031) | 0.068*** (0.020) | 0.058 (0.036) |
| Female Educ, BPL | | | | | | | | | 0.304** (0.134) | | 0.303** (0.134) | | 0.382*** (0.108) | |
| Male Educ, BPL | | | | | | | | | | 0.326** (0.137) | | 0.326** (0.137) | | 0.483 (0.106) |
| Controls | No | No | No | No | No | No | Yes | Yes | No | No | No | No | Yes | Yes |
| Cohort FE | Yes | Yes | Yes | Yes | Yes | Yes | Yes | Yes | Yes | Yes | Yes | Yes | Yes | Yes |
| Metropolitan FE | Yes | Yes | Yes | Yes | Yes | Yes | Yes | Yes | Yes | Yes | Yes | Yes | Yes | Yes |
| Year FE | Yes | Yes | Yes | Yes | Yes | Yes | Yes | Yes | Yes | Yes | Yes | Yes | Yes | Yes |
| Age Quad. | Yes | Yes | Yes | Yes | Yes | Yes | Yes | Yes | Yes | Yes | Yes | Yes | Yes | Yes |
| Observations | 118,113 | 104,337 | 118,113 | 104,337 | 118,113 | 104,337 | 102,029 | 91,036 | 172,433 | 150,353 | 172,433 | 150,353 | 136,268 | 119,281 |

**Notes.** Standard errors, clustered on the country of origin, are reported in parentheses. Controls include the number of siblings, family income, gender inequality index at home country, state-by-year average of unemployment rate, income, and percentage of immigrants. CPS weights are used.

**Table 3.** *Regression Analysis for the effects of average years of schooling in the country of ancestry on educational attainments of first-generation immigrants in different country-groups*

| DV: Education | Low Income | | | | High Income | | | | Comparison | |
|---|---|---|---|---|---|---|---|---|---|---|
| | Female b/se | Male b/se | Female b/se | Male b/se | Female b/se | Male b/se | Female b/se | Male b/se | Female b/se | Male b/se |
| Female Educ, BPL | 0.463*** (0.118) | | | | 0.496*** (0.098) | | | | 0.629*** (0.082) | |
| Low Income × Female Educ, BPL | | | | | | | | | -0.347*** (0.115) | |
| Male Educ, BPL | | 0.491*** (0.165) | | | | 0.491*** (0.121) | | | | 0.755*** (0.120) |
| Low Income × Male Educ, BPL | | | | | | | | | | -0.434*** (0.097) |
| Avg Educ Native | | | 0.055* (0.030) | 0.059** (0.025) | | | 0.065*** (0.020) | 0.039 (0.030) | | |
| Low Income | | | | | | | | | 0.762*** (0.248) | 0.772*** (0.254) |
| Observation | 64854 | 55182 | 64854 | 55182 | 71414 | 64099 | 71414 | 64099 | 136268 | 119281 |

**Notes.** Standard errors, clustered on the country of origin, are reported in parentheses. Controls include the number of siblings, family income, gender inequality index at home country, state-by-year average of unemployment rate, income, and percentage of immigrants. All regressions include fixed effects for the metropolitan area, year, cohort, and a polynomial function of age. CPS weights are used.

Journal of Economics and Political Economy

**Table 4.** *Regression Analysis for the effects of average years of schooling in the country of ancestry on educational attainments of second-generation immigrants in different country-groups*

| | Low Income | | | | High Income | | | | Comparison | |
|---|---|---|---|---|---|---|---|---|---|---|
| | Female | Male | Female | Male | Female | Male | Female | Male | Female | Male |
| DV: Education | b/se | b/se | b/se | b/se | b/se | b/se | b/se | b/se | b/se | b/se |
| Female Educ, MBPL | 0.051 | | | | 0.059*** | | | | | |
| | (0.043) | | | | (0.020) | | | | | |
| Male Educ, FBPL | 0.055*** | | | | 0.064*** | | | | 0.055* | |
| | (0.019) | | | | (0.019) | | | | (0.029) | |
| Low Income × Female Educ, FBPL | | | | | | | | | 0.056 | |
| | | | | | | | | | (0.091) | |
| Male Educ, MBPL | | 0.057 | | | | 0.047** | | | | |
| | | (0.041) | | | | (0.018) | | | | |
| Male Educ, FBPL | | 0.077*** | | | | 0.061*** | | | | 0.030 |
| | | (0.017) | | | | (0.022) | | | | (0.026) |
| Low Income × Male Educ, FBPL | | | | | | | | | | 0.158** |
| | | | | | | | | | | (0.079) |
| Avg Educ Native | | | 0.115*** | 0.089*** | | | 0.127*** | 0.119*** | | |
| | | | (0.021) | (0.024) | | | (0.018) | (0.010) | | |
| Low Income | | | | | | | | | 0.068 | 0.107 |
| | | | | | | | | | (0.064) | (0.065) |
| Observation | 48620 | 43995 | 48620 | 43995 | 75284 | 67126 | 75284 | 67126 | 102029 | 91036 |

**Notes.** Standard errors, clustered on the country of origin, are reported in parentheses. Controls include the number of siblings, family income, gender inequality index at home country, state-by-year average of unemployment rate, income, and percentage of immigrants. All regressions include fixed effects for the metropolitan area, year, cohort, and a polynomial function of age. CPS weights are used.

Table 5. *Regression Analysis for the effects of gender-difference in years of schooling in the country of ancestry on gender-gap in educational attainments of first and second-generation immigrants*

| DV: Education | 2nd Generation | | | | | 1st Generation | | |
|---|---|---|---|---|---|---|---|---|
| | (1) b/se | (2) b/se | (3) b/se | (4) b/se | (5) b/se | (1) b/se | (2) b/se | (3) b/se |
| Sex | -0.036*** (0.014) | -0.036** (0.016) | -0.037** (0.016) | -0.036*** (0.011) | -0.039*** (0.006) | -0.056 (0.041) | -0.056 (0.052) | -0.062* (0.036) |
| Sex* Delta(Educ), MBPL | 0.020 (0.013) | | 0.011* (0.006) | | 0.031*** (0.007) | | | |
| Sex* Delta(Educ), FBPL | | 0.025* (0.013) | 0.020 (0.012) | | 0.025*** (0.006) | | | |
| Sex* Delta(Educ), Natives(2nd) | | | | 0.105*** (0.006) | 0.113*** (0.004) | | 0.106*** (0.018) | 0.136*** (0.024) |
| Sex* Delta(Educ), BPL | | | | | | 0.172*** (0.035) | | 0.192*** (0.031) |
| Delta(Educ), MBPL | -0.044 (0.041) | | -0.019 (0.013) | | -0.050* (0.028) | | | |
| Delta(Educ), FBPL | | -0.068 (0.041) | -0.059 (0.042) | | -0.067*** (0.014) | | | |
| Delta(Educ), BPL | | | | | | -0.455*** (0.112) | | -0.486*** (0.117) |
| Delta(Educ), Natives(2nd) | | | | -0.159*** (0.049) | -0.167*** (0.047) | | -0.191*** (0.058) | -0.232*** (0.066) |
| Cohort Fixed Effects | Yes | Yes | Yes | Yes | Yes | Yes | Yes | Yes |
| Metropolitan Dummies | Yes | Yes | Yes | Yes | Yes | Yes | Yes | Yes |
| Year Dummies | Yes | Yes | Yes | Yes | Yes | Yes | Yes | Yes |
| Age Quad. | Yes | Yes | Yes | Yes | Yes | Yes | Yes | Yes |
| Observation | 193065 | 193065 | 193065 | 193065 | 193065 | 255549 | 255549 | 255549 |

**Notes.** Standard errors, clustered on the country of origin, are reported in parentheses. Controls include the number of siblings, family income, gender inequality index at home country, state-by-year average of unemployment rate, income, and percentage of immigrants. CPS weights are used.

Journal of Economics and Political EconomyTable 6. *Regression Analysis for the effects of gender-difference in years of schooling in the country of ancestry on gender-gap in educational attainments of second-generation immigrants in different country-groups*

| | Low Income | | | High Income | | | Comparison | | | |
|---|---|---|---|---|---|---|---|---|---|---|
| | (1) | (2) | (3) | (1) | (2) | (3) | (1) | (2) | (3) | (4) |
| DV: Education | b/se | b/se | b/se | b/se | b/se | b/se | b/se | b/se | b/se | b/se |
| Sex | -0.012 (0.021) | -0.015 (0.019) | -0.047*** (0.016) | -0.047*** (0.014) | -0.047*** (0.014) | -0.027*** (0.009) | -0.070*** (0.017) | -0.069*** (0.015) | -0.079*** (0.015) | -0.037*** (0.010) |
| Sex* Delta(Educ), FBPL | 0.027 (0.019) | | | 0.026* (0.014) | | | 0.081*** (0.026) | | 0.021** (0.009) | |
| Sex* Delta(Educ), MBPL | | 0.022 (0.018) | | | 0.022 (0.016) | | | 0.072*** (0.024) | 0.060*** (0.023) | |
| Sex* Delta(Educ), Natives(2nd) | | | 0.103*** (0.009) | | | 0.102*** (0.006) | | | | 0.098*** (0.013) |
| Sex* Delta(Educ), FBPL × Low Income | | | | | | | -0.131*** (0.044) | | -0.097*** (0.014) | |
| Sex* Delta(Educ), MBPL × Low Income | | | | | | | | -0.126*** (0.041) | -0.070* (0.037) | |
| Sex* Delta(Educ), Natives(2nd) × Low Income | | | | | | | | | | 0.013 (0.017) |
| Delta(Educ), MBPL | | -0.085 (0.067) | | | -0.020 (0.025) | | | -0.77** (0.038) | -0.071* (0.039) | |
| Delta(Educ), FBPL | -0.125* (0.070) | | | -0.043 (0.032) | | | -0.095** (0.042) | | | |
| Delta(Educ), Natives(2nd) | | | -0.109* (0.065) | | | -0.186*** (0.029) | | | | -0.158*** (0.051) |
| Low Income | | | | | | | -0.148 (0.101) | | -0.112*** (0.031) | -0.038 (0.041) |
| Low Income | | | | | | | | -0.156 (0.104) | -0.110 (0.100) | |
| Observation | 92615 | 92615 | 92615 | 142410 | 142410 | 142410 | 193065 | 193065 | 193065 | 193065 |

**Notes.** Standard errors, clustered on the country of origin, are reported in parentheses. Controls include the number of siblings, family income, gender inequality index at home country, state-by-year average of unemployment rate, income, and percentage of immigrants. All regressions include fixed effects for the metropolitan area, year, cohort, and a polynomial function of age. CPS weights are used.

308

H. NoghaniBehambari, N. Tavassoli, & F. Noghani, 7(4), 2020, p.284-318.

Table 7. *Regression Analysis for the effects of gender-difference in years of schooling in the country of ancestry on gender-gap in educational attainments of first-generation immigrants in different country-groups*

|  | Low Income | | High Income | | Comparison | |
|---|---|---|---|---|---|---|
|  | (1) | (2) | (1) | (2) | (1) | (2) |
| DV: Education | b/se | b/se | b/se | b/se | b/se | b/se |
| Sex | -0.096** | -0.143** | -0.074* | -0.035 | -0.121*** | -0.086** |
|  | (0.039) | (0.066) | (0.038) | (0.024) | (0.036) | (0.041) |
| Sex * Delta(Educ), BPL | 0.140*** |  | 0.117** |  | 0.331*** |  |
|  | (0.035) |  | (0.050) |  | (0.058) |  |
| Sex * Delta(Educ), Natives(1st) |  | 0.156*** |  | 0.076*** |  | 0.113*** |
|  |  | (0.034) |  | (0.012) |  | (0.027) |
| Sex * Delta(Educ), BPL × Low Income |  |  |  |  | -0.249*** |  |
|  |  |  |  |  | (0.077) |  |
| Sex * Delta(Educ), Natives(1st) × Low Income |  |  |  |  |  | -0.026 |
|  |  |  |  |  |  | (0.040) |
| Delta(Educ), BPL | -0.370*** |  | -0.053 |  | -0.392*** |  |
|  | (0.096) |  | (0.176) |  | (0.081) |  |
| Delta(Educ), Natives(1st) |  | -0.333*** |  | -0.088** |  | -0.182*** |
|  |  | (0.054) |  | (0.040) |  | (0.059) |
| Low Income |  |  |  |  | 0.830*** | 0.772** |
|  |  |  |  |  | (0.297) | (0.315) |
| Observations | 120036 | 120036 | 135513 | 135513 | 255549 | 255549 |

**Notes.** Standard errors, clustered on the country of origin, are reported in parentheses. Controls include the number of siblings, family income, gender inequality index at home country, state-by-year average of unemployment rate, income, and percentage of immigrants. All regressions include fixed effects for the metropolitan area, year, cohort, and a polynomial function of age. CPS weights are used.

**Table 8.** *Regression Analysis for the effect of contextual factors in the intergenerational transmission of gender-gap in educational attainments among first-generation immigrants*

|  | 1st Generations | | | |
|---|---|---|---|---|
|  | (1) | (2) | (3) | (4) |
| DV: Education | b/se | b/se | b/se | b/se |
| Sex | -0.062 | -0.063 | -0.078* | -0.062 |
|  | (0.046) | (0.046) | (0.046) | (0.046) |
| Sex * Delta(Educ), BPL | 0.172** | 0.170*** | 0.186*** | -0.179 |
|  | (0.076) | (0.046) | (0.026) | (0.496) |
| Sex * Delta(Educ), BPL × State Avg: Family Income | -0.000 | | | |
|  | (0.001) | | | |
| Sex * Delta(Educ), BPL × State Avg: %Immigrants | | -0.005 | | |
|  | | (0.179) | | |
| Sex * Delta(Educ), BPL × State Avg: Native Education | | | -0.026*** | |
|  | | | (0.008) | |
| Sex * Delta(Educ), BPL × State Avg: Education | | | | 0.026 |
|  | | | | (0.038) |
| Delta(Educ), BPL | -0.406*** | -0.405*** | -0.423*** | -0.408*** |
|  | (0.088) | (0.087) | (0.089) | (0.087) |
| Observations | 291240 | 291240 | 291240 | 291240 |

**Notes.** Standard errors, clustered on the country of origin, are reported in parentheses. Controls include the number of siblings, family income, gender inequality index at home country, state-by-year average of unemployment rate, income, and percentage of immigrants. All regressions include fixed effects for the metropolitan area, year, cohort, and a polynomial function of age. CPS weights are used.

Table 9. *Regression Analysis for the effect of contextual factors in the intergenerational transmission of gender-gap in educational attainments among Second-generation immigrants*

| DV: Education | 2nd Generations | | | | | |
|---|---|---|---|---|---|---|
| | (1) b/se | (2) b/se | (3) b/se | (4) b/se | (5) b/se | (6) b/se |
| Sex | -0.034*** (0.012) | -0.034** (0.014) | -0.034*** (0.012) | -0.034** (0.014) | -0.039*** (0.006) | -0.036*** (0.008) |
| Sex * Delta(Educ), MBPL | 0.067* (0.040) | | 0.031 (0.019) | | 0.036*** (0.005) | |
| Sex * Delta(Educ), MBPL × State Avg: Family Income | -0.001 (0.001) | | | | | |
| Sex * Delta(Educ), FBPL | | 0.083** (0.034) | | 0.035 (0.023) | | 0.036*** (0.006) |
| Sex * Delta(Educ), FBPL × State Avg: Family Income | | -0.001 (0.001) | | | | |
| Sex * Delta(Educ), MBPL × State Avg: %2nd Gen Immigrants | | | -0.127 (0.112) | | | |
| Sex * Delta(Educ), FBPL × State Avg: %2nd Gen Immigrants | | | | -0.106 (0.142) | | |
| Sex * Delta(Educ), MBPL × State Avg: Native Education | | | | | -0.015*** (0.005) | |
| Sex * Delta(Educ), FBPL × State Avg: Native Education | | | | | | -0.013*** (0.005) |
| Delta(Educ), MBPL | -0.036 (0.035) | | -0.037 (0.035) | | -0.067** (0.028) | |
| Delta(Educ), FBPL | | -0.056 (0.037) | | -0.056 (0.037) | | -0.076** (0.031) |
| Observations | 201454 | 201454 | 201454 | 201454 | 201454 | 201454 |

Notes. Standard errors, clustered on the country of origin, are reported in parentheses. Controls include the number of siblings, family income, gender inequality index at home country, state-by-year average of unemployment rate, income, and percentage of immigrants. All regressions include fixed effects for the metropolitan area, year, cohort, and a polynomial function of age. CPS weights are used.

Journal of Economics and Political Economy**Table 10.** *Regression Analysis for the effects of gender-difference in Art and Humanities degrees awarded in the country of ancestry on gender-gap in degrees awarded of first and second-generation immigrants*

|  | First Generations | | | Second and Higher Generations | | |
|---|---|---|---|---|---|---|
|  | (1) | (2) | (3) | (1) | (2) | (3) |
| DV: Art And Humanities | b/se | b/se | b/se | b/se | b/se | b/se |
| Delta(Degree Art, BPL) * Sex | 0.001*** | 0.001*** | 0.001*** |  |  |  |
|  | (0.000) | (0.000) | (0.000) |  |  |  |
| Delta(Degree Art, Ancestry) * Sex |  |  |  | 0.001*** | 0.001*** | 0.001*** |
|  |  |  |  | (0.000) | (0.000) | (0.000) |
| Delta(Degree Art), BPL | 0.004*** | 0.003*** | 0.003*** |  |  |  |
|  | (0.000) | (0.000) | (0.000) |  |  |  |
| Delta(Degree Art), Ancestry |  |  |  | -0.001** | -0.000 | -0.000 |
|  |  |  |  | (0.000) | (0.000) | (0.000)0 |
| Sex (Female = 1) | 0.166*** | 0.152*** | 0.139*** | 0.118*** | 0.141*** | 0.120*** |
|  | (0.005) | (0.004) | (0.006) | (0.006) | (0.006) | (0.006) |
| State Year FE | No | Yes | Yes | No | Yes | Yes |
| State Fixed Effects | No | Yes | Yes | No | Yes | Yes |
| Year Fixed Effects | No | Yes | Yes | No | Yes | Yes |
| Age Quad. | Yes | Yes | Yes | Yes | Yes | Yes |
| Observations | 197787 | 197787 | 197787 | 510772 | 510772 | 510772 |

Notes. Standard errors, clustered on the country of origin, are reported in parentheses. Controls include the number of siblings, family income, gender inequality index at home country, state-by-year average of unemployment rate, income, and percentage of immigrants. Census weights are used.

312

H. NoghaniBehambari, N. Tavassoli, & F. Noghani, 7(4), 2020, p.284-318.

## A2. Figures

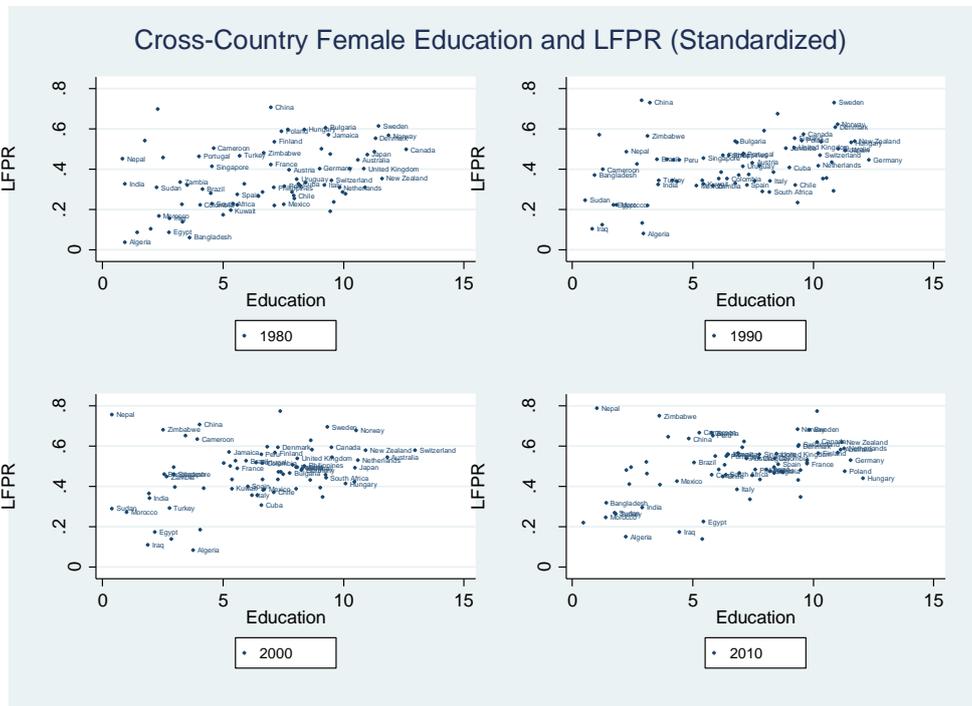

**Figure 1.** *Standardized Annual Average Levels of Education and Labor Force Participation Rates of Women Across the Glob*

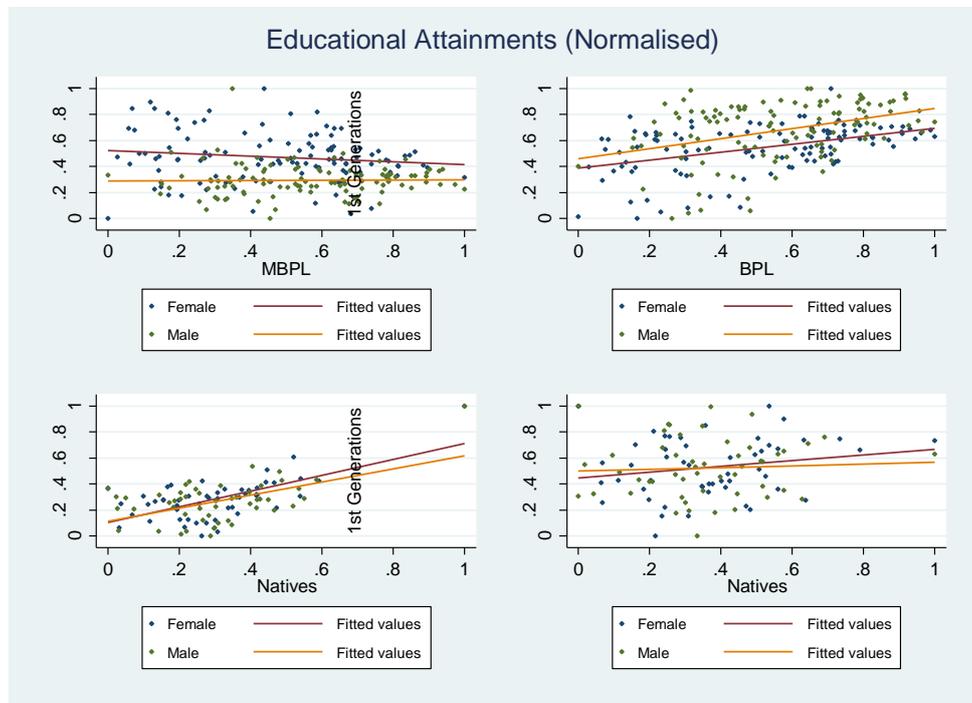

**Figure 2.** *Educational Attainments (Levels, Normalized)*

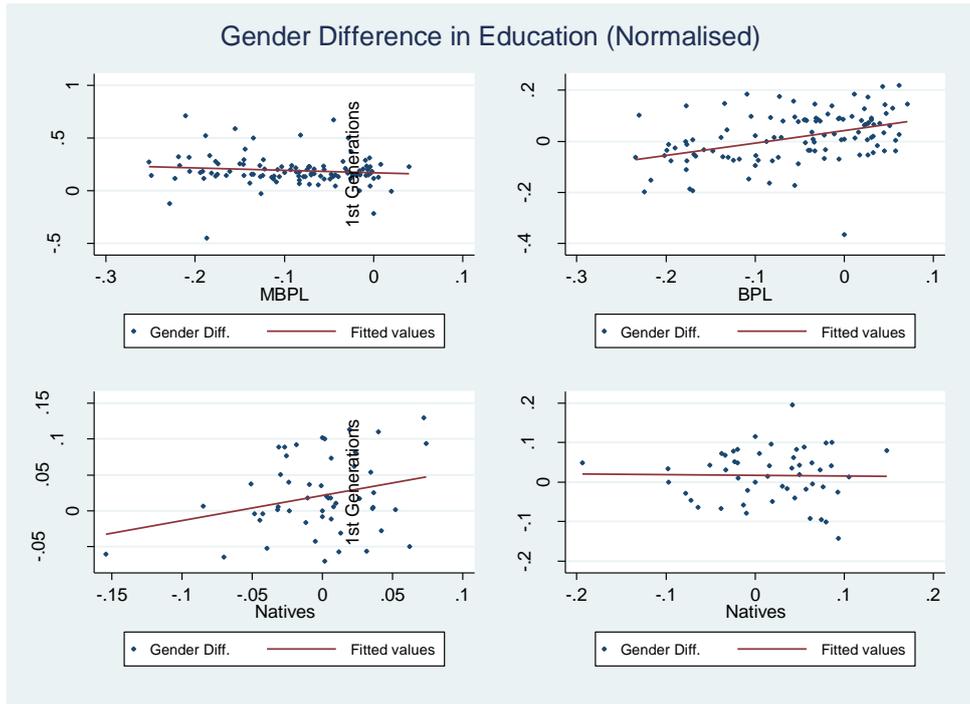

**Figure 3.** *Gender Gap in Educational Attainments (Female-Male Di, Normalized)*

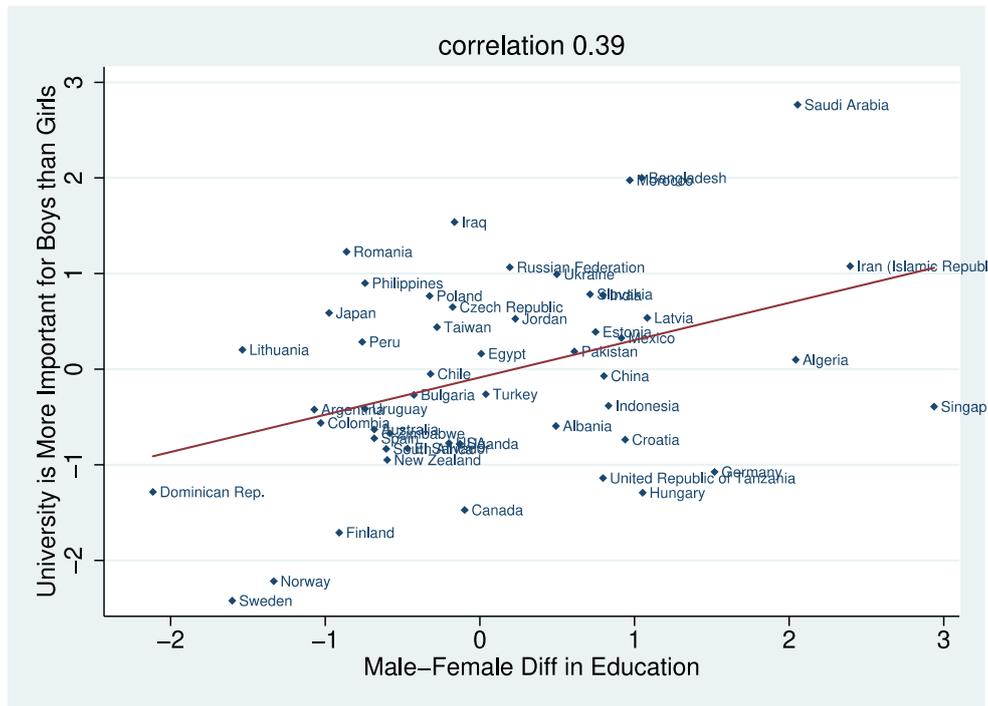

**Figure 4.** *Standardized response to the question that whether the university is more important for boys than girls. The data is extracted from Waves 3 and 4 of the World Value Survey dataset (covering the years 1994-1998 and 1999-2004). Gender differences in education are calculated as extra years of schooling of males compared to females. The data is extracted from Barro-Lee dataset. The time-span of the latter dataset is restricted to the year 2000. Standardized differences are linked to the WVS dataset.*

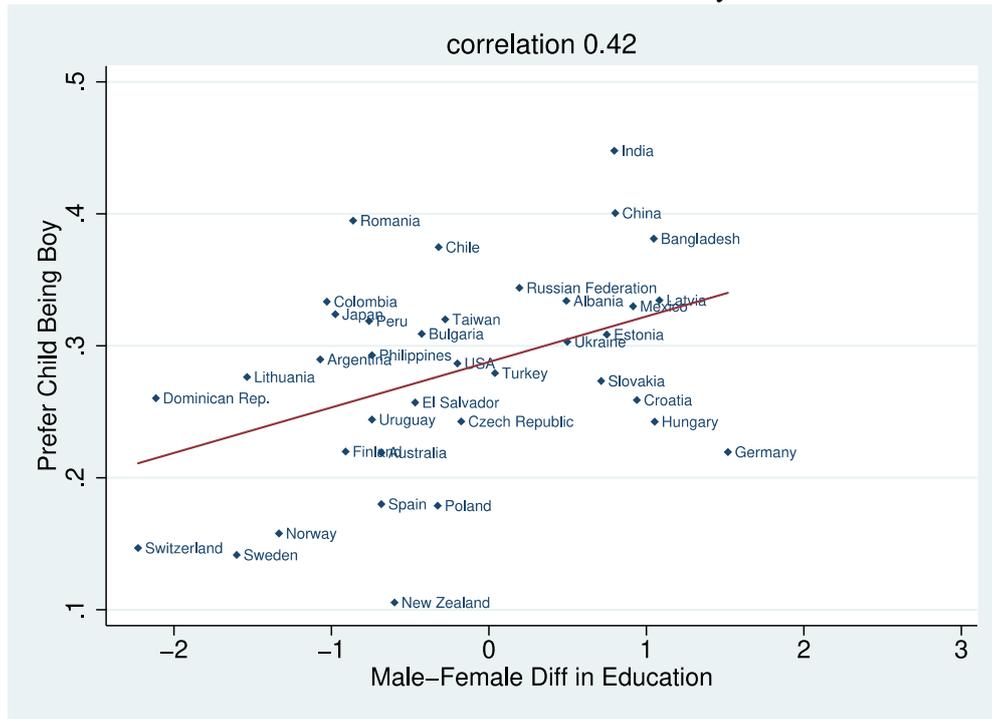

**Figure 5.** *Standardized response to the question that could you have only one child would you prefer a boy or a girl. The data is extracted from Wave 3 of the World Value Survey dataset (covering the years 1994-1998). Gender differences in education are calculated as extra years of schooling of males compared to females. The data is extracted from the Barro-Lee dataset. The time-span of the latter dataset is restricted to the year 2000. Standardized differences are linked to the WVS dataset.*

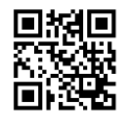